\documentclass[prd,12pt,nofootinbib,tightenlines]{revtex4}
\usepackage{graphicx}

\usepackage{amsmath}

\newcommand{\lepto}{{\sc Lepto}}
\newcommand{\pythia}{{\sc Pythia}}

\begin{document}

\title{Colour rearrangements in $B$-meson decays}
\author{David Eriksson}
\email{david.eriksson@physics.uu.se}
\author{Gunnar Ingelman}
\email{gunnar.ingelman@physics.uu.se}
\author{Johan Rathsman}
\email{johan.rathsman@physics.uu.se}
\affiliation{High Energy Physics, Department of Physics and Astronomy, Uppsala University
Box 535, SE-75121 Uppsala, Sweden}

\begin{abstract}
We present a new model, based on colour rearrangements, which at the same time can describe both hidden and open charm production in $B$-meson decays. The model is successfully compared to both inclusive decays, such as $ B \to J/\psi X$ and $ B \to D_s X$, as well as exclusive ones, such as $ B \to J/\psi K^{(\ast)}$ and $ B \to D^{(\ast)} D^{(\ast)}K $. It also gives a good description of the momentum distribution of direct $J/\psi$'s, especially in the low-momentum region, which earlier has been claimed as a possible signal for new exotic states.
\end{abstract}

\maketitle

\section{Introduction}

A proper understanding of the confinement phenomenon in quantum
chromodynamincs (QCD), describing the transition from the perturbatively 
calculable parton level to the experimentally observable hadron level, 
is still missing.
In order to get a better understanding of the hadronisation process one is
therefore lead to constructing models, such as the Lund string fragmentation 
model~\cite{Lundstring}, and then compare these to data. 
Of special interest in these types of models is the treatment of the colour 
quantum number and the associated colour flux. 
Typically, the planar approximation is used for this, which is valid in the 
large $N_C \to \infty$ limit, leading to a good description of inclusive
event properties at high energy colliders. 

At the same time, there is also a large class of so called hard diffractive 
processes in both $ep$ and $pp$ collisions which cannot be described by
the planar approximation. 
The signifying feature of this type of processes is that the final state 
particles are divided into two or more colour singlet systems, separated 
by large rapidity gaps, and that at least one of these systems has the 
properties of a hard partonic interaction, such as jets. 
These events are in accordance with the predictions of the model introduced \cite{Ingelman:1984ns} based on pomeron exchange. Although further developments of this model is successful in describing rapidity gap data, it implies different descriptions of diffractive and non-diffractive events without a smooth transition in between.

In order to order to remedy this situation and get a model that can describe
both inclusive and hard diffractive processes, the Soft Colour Interaction 
(SCI) model was introduced~\cite{SCI}. 
In short, this model is based on the assumption that the colour flux from the
hard perturbative interaction is modified through interactions with the
colour background field represented by the remnants of the incoming hadrons.
In the simplest version of the model there is only one new additional parameter
describing this interaction, namely the probability for such a colour exchange.
These colour rearrangements were added to the Lund Monte Carlo programs \lepto~\cite{Ingelman:1996mq} for deep inelastic scattering (DIS) and \pythia~\cite{pythia} for hadron-hadron collisions, where they give rise to events having regions in phase space where no string is stretched, and therefore no hadrons are being produced. These rapidity gap events are classified as diffractive and this simple model essentially reproduces all data on diffractive hard scattering in both 
$ep$~\cite{SCI} and $pp$~\cite{Enberg:2001vq} collisions. Of course, events without such rapidity gaps are also produced corresponding to ordinary inclusive events. 
The SCI colour rearrangements can also turn a colour octet $c\bar{c}$ or $b\bar{b}$ pair into a singlet, leading to production of charmonium and bottomonium states in basic agreement with data from $pp$ \cite{Edin:1997zb} and $pA$ and $\pi A$ \cite{BrennerMariotto:2001sv}.

This phenomenological success of the SCI model \cite{Ingelman:2005ku} indicates that it captures some very essential QCD dynamics. It is therefore interesting that recent developments on QCD rescattering theory \cite{Brodsky:2002ue} provides a basis for this model \cite{Brodsky:2004hi}. Rescatterings of a hard-scattered parton on the spectator system cannot be gauged away and do contribute at leading twist. In DIS such rescatterings of the struck quark via 1,2\ldots gluons are summed in the Wilson line used in the definition of the parton density functions, which thereby absorb these rescatterings effects when fitted to inclusive DIS data. However, for less inclusive observables that depend on the colour structure in the event, these rescatterings are important and the SCI model is a phenomenological model to account for their effect.

There are also several extensions of the model.  The  difference in the
potential energy of various string configurations  can be included~\cite{GAL}
and the momentum transfer in the colour exchange  can be
modelled~\cite{BrennerMariotto:2001sv}.  The model has also been successfully
extended to describe jet quenching in a  quark gluon plasma~\cite{Zapp:2005kt}.

The rational behind the SCI model is to learn more about the non-perturbative
dynamics by starting from a well defined perturbative state. One such example,
which is the subject of this paper, is the production of hidden and open charm
in hadronic $B$-decays. Thanks to the $B$-factories there is now a wealth of
detailed data from the BaBar and Belle experiments, which can be used as a
testing ground for the ideas behind the SCI model. 
In a $B$-meson decay the hard scale is given by the $b$-quark mass, 
$m_b \approx 5 $ GeV, and the decay products then interact with the remaining
soft part of the $B$-meson. Following earlier
applications of the SCI model, we are aiming at formulating a model which can
describe both hidden and open charm production.

This paper is organized as follows. We start in section 2 by reviewing earlier
models for hidden and open charm production in B-decays, with emphasis on
charmonium production, and how they compare with data. 
In section 3 we then present our model on the parton level and the
transition to the observable hadron level. The resulting model is
then compared to existing data in section 4 and finally section 5 contains the conclusions.

\section{Earlier models}

Charmonium production in $B$-decays has a long history. In a naive version of
the so called colour singlet model (CSM)~\cite{DeGrand:1979wf} one notes that there is a 
probability 1/9 that the $c\bar{c}$ pair in a $b\to c\bar{c}s$ decay is 
in a colour singlet state and combines this with the $J/\psi$ wave-function 
at the origin giving a rate which is in reasonable agreement with data.
However, in a proper treatment one also has to take into account that the
Hamiltonian describing the $b$-decay should be the effective one, where the
$W$-boson (and $t$-quark) has been integrated out~\cite{Wise:1979tp}. 
This gives an additional factor $\sim 15$ suppression of the colour 
singlet state compared to the colour octet one~\cite{Beneke:1998ks}. 
To make things even worse, a strict NLO calculation of the colour singlet rate 
becomes negative~\cite{Bergstrom:1994vc} unless one also includes 
${\cal O}(\alpha_s^2)$ corrections to the colour octet channel and thus modifies
the perturbative expansion. Even so, the rate obtained is about a factor ten below data. 
In addition the CSM cannot be used to calculate the production of P-wave
charmonium states ($\chi_c$). 

A theoretically more sound description of charmonium production in
$B$-decays is provided by the so called colour octet model (COM), based on
non-relativistic QCD~\cite{NRQCD}, in which the decay is factorised into two parts. 
First, a non-relativistic colour singlet or octet $c\bar{c}$-pair is produced 
in a given spectroscopic state, $^{2S+1}L_J$, where $S$, $L$ and $J$ are 
the spin, orbital angular momentum and total angular momentum respectively, 
and then this state is transformed into a charmonium hadron through the 
possible emission of soft gluons in order to get a colour singlet system. The
latter process is described by non-perturbative matrix-elements which in
principle can be fitted from data. 

If one only takes into account the colour singlet and octet $^{3}S_1$ states
and uses data from the $J/\psi$ decay width and the production of $J/\psi$
at large transverse momenta at the Tevatron to fit the non-perturbative matrix
elements, then the COM predicts a rate for direct $J/\psi$ production which is
about half of the observed value~\cite{Ko:1995iv} and similarly for $\psi^\prime$.
However, in a proper non-relativistic expansion the colour octet 
$^{1}S_0$ and $^{3}P_J$, $J=0,1,2$ states also need to be 
included~\cite{Beneke:1998ks}.
Unfortunately, the corresponding non-perturbative matrix-elements
cannot be fitted independently so it is not possible 
to get a prediction for these rates. Once the $J/\psi$ rate has been fitted 
it can be used~\cite{Beneke:1998ks} to get a prediction for the inclusive $\eta_c$ 
rate, for which there is yet no data. 
Finally one can fit the branching ratio $B \to \chi_{c2} X$ and get a 
prediction for $B \to \chi_{c1} X$ which turns out to be about a factor 
two below the observation. 

Given that the $J/\psi$ rate is fitted to data, the COM can then be used to
calculate the  $J/\psi$ momentum distribution by assuming that the $b$-quark 
decays through a two body process $b \to (c\bar{c}) q$, taking into account 
smearing from the boost to the $B$ (or $\Upsilon(4S)$) rest 
system~\cite{Palmer:1996dy} as well as a non-perturbative ``shape-function" which 
resums the emissions of multiple soft gluons~\cite{NRQCD-mom}. 
Overall this gives a good description of the data on direct $J/\psi$ production when
combined with a model for the $J/\psi K(^\ast)$ contributions except for small 
momenta where it is much below the data~\cite{BaBarJpsi}. Several suggestions 
have been made for how to describe this low-momentum region, including enhanced
baryon pair production~\cite{Brodsky:1997yr}, hybrid 
mesons~\cite{Close:2003ae, Chua:2003fp}, and
hidden charm diquark-antidiquark bound states~\cite{Bigi:2005fr}. 

As already alluded to our model aims to describe both open and hidden charm
production. Whereas the production of charmonium states in $B$-decays have been
studied extensively, there has been less attention devoted to 
$B \to DD_s$ and $B \to DDK$ 
decays. There are some models, which all are more or less 
based on the factorization hypothesis and to varying degree make use of 
heavy quark symmetry. For example,  for $B \to DD_s$ decays
we have the simpler 
pole models~\cite{Bauer:1986bm,Rosner:1990xx} which later have been 
refined~\cite{Luo:2001mc,Chen:2005rp} using heavy quark symmetry.  
Another example  is given by~\cite{Thomas:2005bu},
which is an extension 
of the so called Isgur-Scora-Grinstein-Wise model~\cite{Isgur:1988gb} 
to non-leptonic decays, also assuming factorisation. 
Typically all of these models are able to describe the branching ratios of the
specific decay processes they are studying
(see for example~\cite{Thomas:2005bu}) but they cannot be used to 
predict hidden charm production. 
Finally, there also exists models for $B \to DDK$ decays assuming that they
proceed via an  intermediate $DD_{sJ}$ state~\cite{Colangelo:2002dg,Datta:2003re}, which
however are not so successful at describing data.

\section{Our model}

The model has three main ingredients: The internal $B$-meson dynamics with
the $b$-quark decay giving a partonic final state, the soft colour interactions
which modifies the colour structure of the event and  finally hadronisation
using the Lund string model amended with special treatment of systems of small
invariant mass and the mapping onto discrete charmonium states. It is
particularly important to have a properly devised and tuned hadronisation model
in order to describe the exclusive few-body final states in $B$-meson decays
that is investigated here. In this section we define and discuss these details of
our model. 

\subsection{$B$-meson dynamics and $b$-quark decay}

The $B$-meson can be viewed as a non-relativistic $b$-quark surrounded by a non-perturbative hadronic system represented by a spectator quark to account for the light valence quark as well as seaquarks and gluon contributions. This is the basis for the ACCMM model~\cite{ACCMM}, which describes the internal dynamics of the $B$-meson. In the $B$-meson rest system the three-momentum $\mathbf{p}_b$ of the $b$-quark is spherically symmetric and given by the normalised Gaussian
\begin{equation}
\begin{aligned}
\Phi(|\mathbf{p}_b|) \; = \; 
\frac{4}{\sqrt{\pi} p_F^3}e^{{-\mathbf{p}_b^2/p_F^2}},
\end{aligned}
\label{eq:b-mom}
\end{equation}
having a width given by the parameter $p_F$. 
The other model parameter is the mass $m_{sp}$ of the spectator. The
decaying $b$-quark is not a final state parton and its mass $m_b$ is allowed to
vary dynamically as given by energy-momentum conservation,
$M_B^2=(p_b+p_{sp})^2$, resulting in
\begin{equation}
\begin{aligned}
m_b^2 \; = \; M_B^2 + m_{sp}^2 - 2 M_B \sqrt{\mathbf{p}_b^2 + m_{sp}^2}\; .
\end{aligned}
\label{eq:b-mass}
\end{equation}

\begin{figure}
\includegraphics*{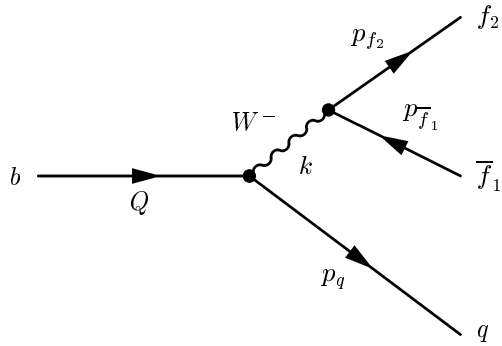}
\caption{Feynman diagram of the weak $b$-quark decay into a lighter quark, $q$, and a fermion pair, $\bar{f}_1$ and $f_2$. $Q$ is the momentum of the incoming $b$-quark, $k$ is the momentum transferred by the $W$ and $p$ the momenta of final state particles.}
\label{feynbdec}
\end{figure}

The weak decay of the $b$-quark is illustrated in Fig.~\ref{feynbdec}, which
defines the four-momenta in the process, where in the $b$ rest frame
$Q=(m_b,\mathbf{0})$. The differential decay rate in the $b$ rest frame is then
\begin{widetext}
\begin{equation}
\begin{aligned}
d\Gamma_{\rm h/sl}\;=\;&
K_{\rm h/sl} \left(2\pi\right)^4 \delta^4(Q  -p_q  - p_{\bar{f}_1} - p_{f_2}) 
\frac{1}{2m_b} 
 \frac{1}{2} \sum_{spins} |\mathcal{M}|^2 \frac{d^3 p_q}{2E_q(2\pi)^3} 
 \frac{d^3 p_{\bar{f}_1}}{2E_{\bar{f}_1}(2\pi)^3} 
 \frac{d^3 p_{f_2}}{2E_{f_2}(2\pi)^3} ,
\label{eqndiffdec}
\end{aligned}
\end{equation}
where $K_{\rm h/sl}$ are $K$-factors for the hadronic and semi-leptonic decays, respectively, that will be fitted to data to get the correct normalization. 
The leading order spin averaged squared matrix element for the $b$-quark 
decay can be expressed as \cite{Collider}
\begin{equation}
\begin{aligned}
\frac{1}{2} \sum_{spins} |\mathcal{M}|^2\;=\;&
64 G_F^2 |V_{f_2 f_1}|^2 |V_{q b}|^2 (p_b \cdot p_{f_2})
(p_q \cdot p_{\bar{f}_1})
\frac{M_W^4}{(k^2 - M_W^2)^2 + \Gamma_W^2 M_W^2}
\label{eqnmatel}
\end{aligned}
\end{equation}
\end{widetext}

In our Monte Carlo model we thus start by generating the three-momentum $\mathbf{p}_b$ of the $b$-quark from Eq.~(\ref{eq:b-mom}), which is then used in Eq.~(\ref{eq:b-mass}) to get the dynamical $b$-quark mass $m_b$. The momenta of the decay products from the $b$-quark decay are then generated according to the differential decay rate in Eqs.~(\ref{eqndiffdec},\ref{eqnmatel}) in the $b$-quark rest
frame and boosted to the $B$-meson rest frame.

\subsection{Colour structure}
\label{sec:colstr}

\begin{figure*}
\begin{tabular}{|@{\hspace{25pt}}c@{\hspace{25pt}}|@{\hspace{25pt}}c@{\hspace{25pt}}|} \hline
\parbox[b]{0pt}{
\setlength{\unitlength}{10mm}
\begin{picture}(0.8,1)
\linethickness{0.4mm}
\put(.4,.4){\circle{0.7}}
\put(0,0){\makebox(0.8,0.8){\large{$\mathbf{I}$}}}
\end{picture}

\rule[-20pt]{0pt}{100pt}}
\includegraphics{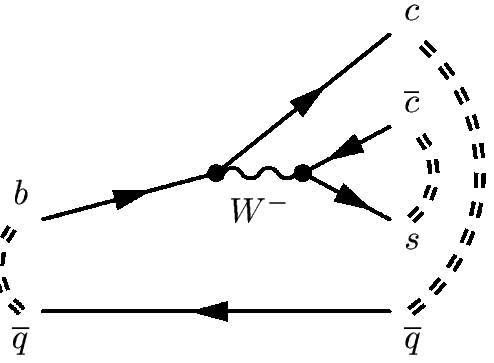} 
&
\parbox[b]{0pt}{
\setlength{\unitlength}{10mm}
\begin{picture}(0.8,1)
\linethickness{0.4mm}
\put(.4,.4){\circle{0.7}}
\put(0,0){\makebox(0.8,0.8){\large{$\mathbf{I\!I}$}}}
\end{picture}

\rule[-20pt]{0pt}{100pt}}
\includegraphics{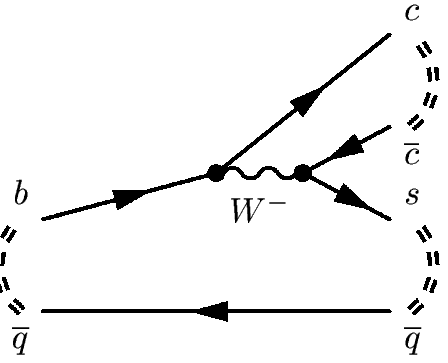}
\\ \hline
\end{tabular}
\caption{Feynman diagram showing the colour string connection with normal colour connection ($\mathbf{I}$) and with reconnected colour ($\mathbf{I\!I}$). The double dashed lines indicate the colour string connections.}
\label{feynncrc}
\end{figure*}

From the calculation of the two colour configurations in the $b\to c\bar{c}s$ 
decay using the effective theory~\cite{Wise:1979tp}, we know that the $c\bar{c}$ colour singlet fraction is suppressed with about a factor 100 compared to the colour octet one. 
Therefore, we will in the following simply assume that all the parton level
decay products from the $B$-meson are in the colour configuration represented by
diagram $\mathbf{I}$ in Fig.~\ref{feynncrc}, where $c\bar{q}$ is one colour
singlet system and  $\bar{c}s$ the other as indicated in the figure.

In order to model the colour suppressed mode where the  $c\bar{c}$-pair forms a
colour singlet, as illustrated in diagram $\mathbf{I\!I}$ in Fig.~\ref{feynncrc}, 
we use the Soft Colour Interaction (SCI) model~\cite{SCI}. This model is based on the assumption that partons emerging from some hard, perturbative process interact softly with the colour background field provided by the spectators of the initial hadron as they propagate through it on a Fermi length scale. In these soft processes, the small momentum transfers are not important and can be neglected, at least in a first approximation. Instead, it is the colour exchange that is important, since it changes the colour string topology of the event and thereby affects the hadronisation giving a different hadronic final state. The model uses an explicit mechanism where colour-anticolour, corresponding to a non-perturbative gluon, can be exchanged between partons and remnants. This should be a natural part of the process when bare partons are dressed into non-perturbative ones and the confining colour flux tubes (strings) is formed between them. 

The SCI model has been added to the Lund Monte Carlo programs \lepto~\cite{Ingelman:1996mq}
 for deep inelastic scattering (DIS) and \pythia~\cite{pythia} and for hadron-hadron collisions. The hard parton level interactions are given by standard perturbative matrix elements and parton showers, which are not altered by the softer non-perturbative effects occurring on a longer space-time scale. The probability for a SCI in terms of the exchange a soft gluon within any pair of a parton and a spectator remnant cannot be calculated and is therefore taken as a constant given by a phenomenological parameter $P$, which is the only parameter of the model. 
As mentioned in the Introduction, this model has support from QCD rescattering theory and is very successful in describing data on hard diffraction, i.e.\ rapidity gap events, and charmonium production. 

Applying the SCI model on $B$-meson decays means that the perturbative partons
from the $b$-quark decay will undergo soft non-perturbative interactions with
the background colour field in the $B$-meson represented by the spectator quark.
Again, the parameter $P$ specifies the probability for such a colour-anticolour
exchange between any of the partons with the spectator. Since there are here
only 4 partons, including the spectator, there are only two possible string
configurations\footnote{Due to the small phase space for extra emissions the 
parton shower does not give any additional gluons and strings for a large 
majority of decays.} as shown in Fig.~\ref{feynncrc}. Starting from 
configuration $\mathbf{I}$, a soft gluon exchange switches the system to 
configuration $\mathbf{I\!I}$. A second such colour exchange switches the system
back to $\mathbf{I}$ again. Thus, increasing the colour exchange probability 
$P$ too much will not favour configuration $\mathbf{I\!I}$. This
switch-back effect is extreme in this 4-parton state. A similar effect appears
for the rapidity gap rate in high energy $ep$ and $pp$ collisions, where
additional colour exchanges may switch back from a gap topology to a no-gap
topology, but is less pronounced due to the presence of more partons. As shown
below, the fit to $B$-meson decays gives the value $P\approx 0.15$, whereas the
rapidity gap rate is not strongly dependent on $P$ and the model is stable for
$P\simeq 0.2-0.5$ \cite{Enberg:2001vq}.

\subsection{Hadronisation of small mass string-systems}
\label{smallen}

In order to be able to compare our model with data we also need to describe the
transition from the parton level to hadrons. As starting point we use the Lund
string model \cite{Lundstring} as implemented in the \pythia\ Monte Carlo \cite{pythia} together with its special
treatment of string systems with small invariant masses, which produce only one or two hadrons. For $B$-meson decays
the latter part of the model is most important and, as we will see below, we have
had to introduce a more careful treatment of these small-mass systems 
in order to describe data on exclusive decay modes. In addition, to calculate the probabilities for different charmonium states we use the model~\cite{BrennerMariotto:2001sv}, as will be discussed in the next subsection.

In the standard Lund string fragmentation picture hadrons are produced
iteratively by considering pair production of quark-antiquark pairs\footnote{In order to be brief we only describe here the production of mesons. The Lund string fragmentation model can also describe the production of baryons, for example by production of diquark-antidiquark pairs.} in the
colour field of the string leading to the production of two colour singlet
systems: one of which is a quark-antiquark pair which becomes a
meson and the other which is a rest string. This
process is then repeated until the rest string has such a low invariant mass
that the procedure is terminated by producing either two mesons or a single
one as will be discussed below.

Normally, for a given quark anti-quark pair that is going to form a meson, the
model in \pythia\ only produces, i.e.\ maps the pair onto, the two lowest order
mesons, i.e.\ those with $L=0,\: S=0,\: J=0$ (such as $K$-mesons) and 
$L=0,\: S=1,\: J=1$ (such as $K^\ast(892)$) mesons.
However, to be able to fit both inclusive, where all type of mesons contribute, and exclusive branching ratios, where mostly the lowest order mesons contribute,
we also need to activate the production of Axial Vector Mesons (AVM) with
$L=1, \: S=0,\: J=1$ (such as $K_1(1270)$). The production of these mesons
are controlled by the \pythia -parameter $P_{\rm AVM}$ ({\tt PARJ(14)}),  which is
the probability that a $S=0$ meson is in a $L=1$ state. These mesons have
substantially higher mass than the ordinary ones, which necessitates refinements
to the model for their production in small mass systems.

Consider a low mass system, with invariant mass $m_{sys}$.  To hadronise this
system \pythia\ first tries $n_{try}$ ({\tt MSTJ(17)}) times to make two hadrons
with total mass below $m_{sys}$. The two hadrons are given a relative momentum to
conserve the invariant mass of the system. If the program fails to make two
hadrons, one hadron is made instead and put on-shell by exchanging an effective gluon
with some other part of the event. We have changed this second step to also
try $n_{try}$ times to make a single hadron with mass below $m_{sys}$ and if this
fails accept the last tried hadron even though $m_h>m_{sys}$.

In the one hadron case, there are several ways to put the hadron
on-shell and obey energy-momentum conservation.
In default \pythia\ it is done in different ways depending on the situation: If
the system in question is the only one left to hadronise, it exchanges momentum
with the, already produced, final state particle which is furthest away in
momentum space. When another unhadronised system exists two different
procedures are used. If $m_{had} < m_{sys}$, the four-momentum vector,
$p_{had}$, is scaled down and the excess momentum is put as a gluon in the other
system. If $m_{had} > m_{sys}$, the four-momentum needed is taken from the other
system.

\begin{figure*}
\begin{footnotesize}
\includegraphics*[width=16cm]{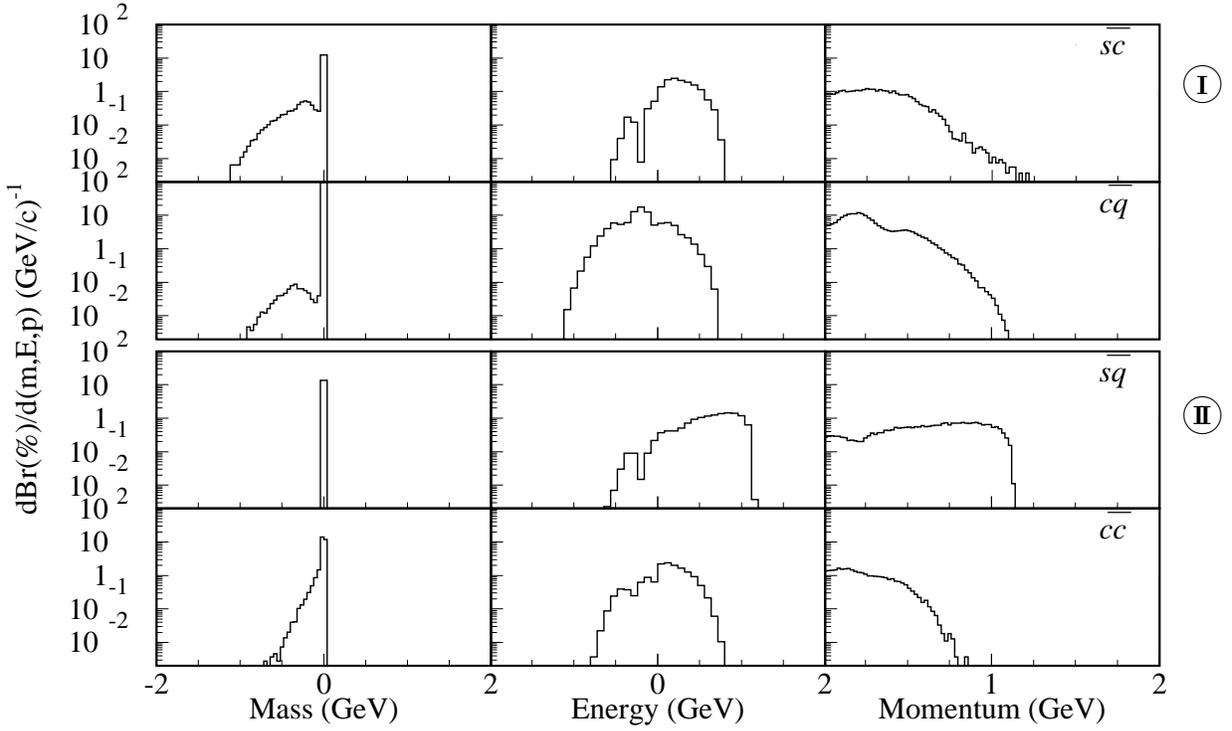}
\setlength{\unitlength}{10mm}
\begin{picture}(-0.2,7)
\linethickness{0.4mm}
\put(-0.2,8.8){\circle{0.5}}
\put(-0.6,8.4){\makebox(0.8,0.8){{$\mathbf{I}$}}}
\put(-0.2,4.4){\circle{0.5}}
\put(-0.6,4.0){\makebox(0.8,0.8){{$\mathbf{I\!I}$}}}
\end{picture}
\end{footnotesize}
\caption{Distribution in mass, energy and three-momentum for the effective
gluon, with $p^2=0$, which is transferred from the specified $q
\bar{q}$-pair, when it hadronises into only one particle. Indicated in the figure are the color connections for the pair, see Fig.~\ref{feynncrc}.} \label{gluon}
\end{figure*}

We have tried a number of different detailed treatments for momentum
exchange to unhadronised systems. It turns out that this part of the model has
little influence on the branching ratios, but can affect the $J/\psi$-momentum
distribution.
The best result is obtained when requiring that the effective gluon has
$p_g^2=0$.
 In practice this is done by giving the gluon a fraction $x$ of the system
three momentum and $(1-x)$ to the hadron, i.e.
$p_g=(x|\mathbf{p}_{sys}|,x\mathbf{p}_{sys})$ and
$p_{had}=(E_{sys}-x|\mathbf{p}_{sys}|,(1-x)\mathbf{p}_{sys})$, with the
condition $p_{had}^2=m_{had}^2$.
Fig.~\ref{gluon} shows the mass, energy and momentum distributions of the
exchanged gluon for the different parton system cases. It can be noted
that $p^2$ is not always zero,
because if there are no unhadronised systems to 
exchange momentum with then
momentum is instead exchanged with another 
final state particle as described above.

\subsection{Hadronisation to charmonium states}
A special case of hadronisation of small mass systems is the mapping of colour
singlet $c\bar{c}$ systems produced at the parton level with a continuous mass
spectrum onto the discrete mass spectrum of charmonium states. To calculate
the probability to obtain different charmonium states we use the 
model~\cite{BrennerMariotto:2001sv}, which is based on the assumption
that it is more likely that a $c \bar{c}$\ pair of given mass $m_{c \bar{c}}$ is
mapped to a charmonium state which is close in mass rather than to one
which is
further away. This can be motivated by the fact that the charmonium mass
spectrum covers a mass range of almost 1 GeV, which is substantially larger than
the few hundred MeV energy-momentum transfers of the soft colour interactions
that may affect the invariant mass of the system. For example, a $c\bar{c}$ with
mass just above the threshold $2m_c$, using $m_c=1.35$ GeV, should have a
larger probability to produce a $J/\psi$ than a $\psi^\prime$, and a $c\bar{c}$
close to the open charm threshold should contribute more to $\psi^\prime$ than
to $J/\psi$. 

Thus, the model assumes that the smearing of the $c \bar{c}$ mass due to soft
interactions is described by the Gaussian
\begin{equation}
\begin{aligned}
G_{sme}(m_{c \bar{c}},m) = \exp \left(-\frac{(m_{c \bar{c}} - m)^2}{2 \sigma_{sme}^2}\right),
\end{aligned}
\end{equation}
where the width is $\sigma_{sme}=0.4$ GeV.
The probability  that a $c\bar{c}$ pair of mass $m_{c \bar{c}}$ forms a
charmonium state $i$ of mass $m_i$ is then given by
\begin{equation}
\begin{aligned}
P_i(m_{c \bar{c}}) = 
\frac{\int G_{sme}(m_{c \bar{c}},m) F_i(m_i,m)dm}
     {\sum_j \int G_{sme}(m_{c \bar{c}},m) F_j(m_j,m)dm},
\end{aligned}
\end{equation}
where $F_i(m_i,m)$ is the distribution in invariant mass for a given  charmonium
state $i$. For our purposes it is enough to use the approximation $F_i(m_i,m) =
s_i \delta(m-m_i)$, i.e.\ neglecting the very narrow width of charmonium states
but including the relative weights $s_i = 2J_i + 1$ coming from non-relativistic
spin statistics.\footnote{We do not need to include any additional
suppression factor $1/n$ for states with higher main quantum number $n$, which
was included in the earlier studies \cite{Edin:1997zb,BrennerMariotto:2001sv}, 
since the model used then did not include the additional suppression of heavy
mesons from trying more than once to make a single hadron with 
$m_{had} < m_{sys}$.} The expression we use is then
\begin{equation}
\begin{aligned}
P_i(m_{c \bar{c}}) = 
\frac{s_i G_{sme}(m_{c \bar{c}},m_i)}
     {\sum_j s_j G_{sme}(m_{c \bar{c}},m_j)}.
\end{aligned}
\end{equation}

\begin{figure*}
\includegraphics*[width=16cm]{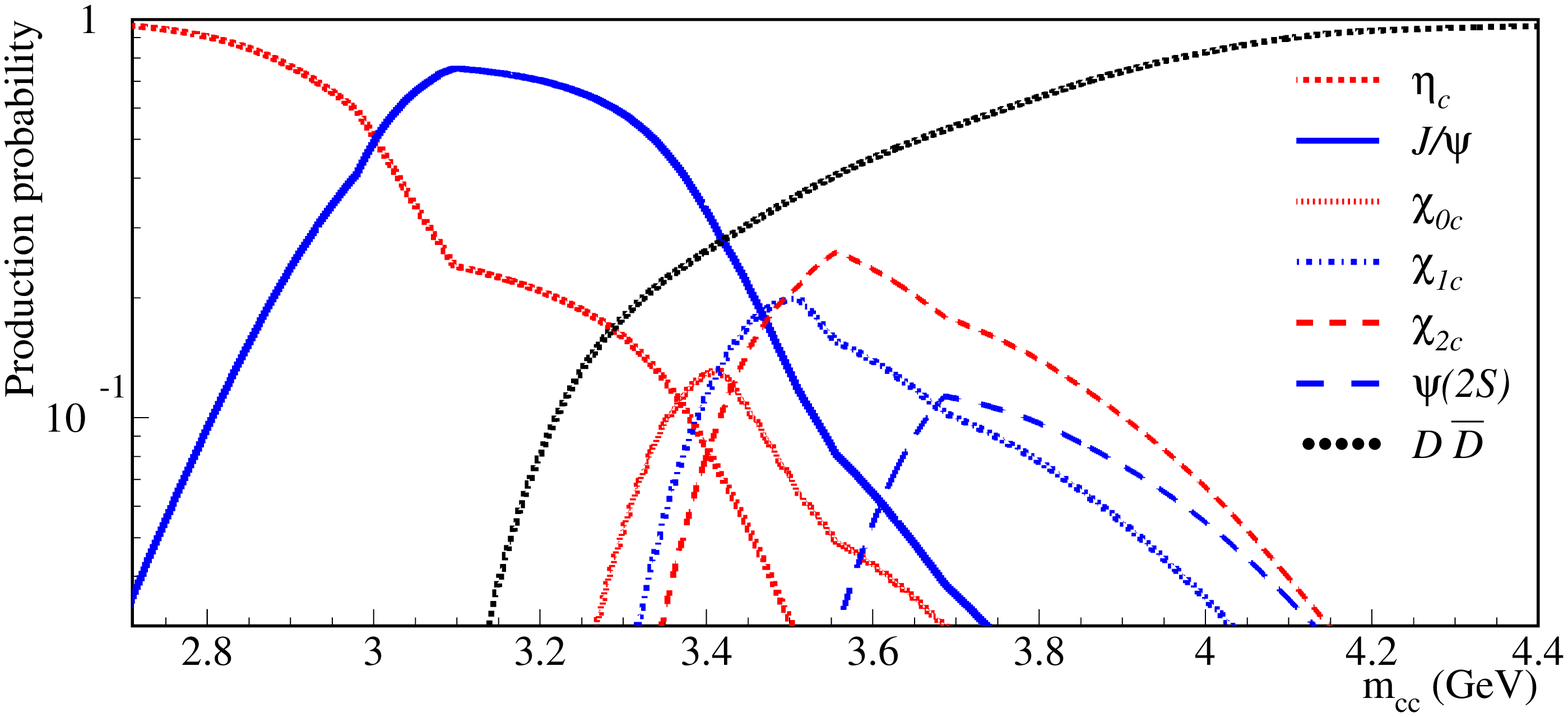}
\caption{Probabilities used for mapping $c \bar{c}$ pairs of mass $m_{c\bar{c}}$ onto different charmonium states and open charm states ($D \bar{D}X$).}
\label{ccbarprob}
\end{figure*}

In our study the following six charmonium states have been included, 
$\eta_c$, $J/\psi$, $\chi_{0c}$, $\chi_{1c}$, $\chi_{2c}$ and $\psi(2S)$.
Figure~\ref{ccbarprob} shows the resulting probabilities for the different charmonium states. It includes the decrease in total probability for charmonium production due to open charm production and the effect from trying more than once to make a single hadron giving an additional suppression of heavy mesons at low $m_{c\bar{c}}$.

\section{Results}

\begin{figure*}
\begin{footnotesize}
\includegraphics*[width=16cm]{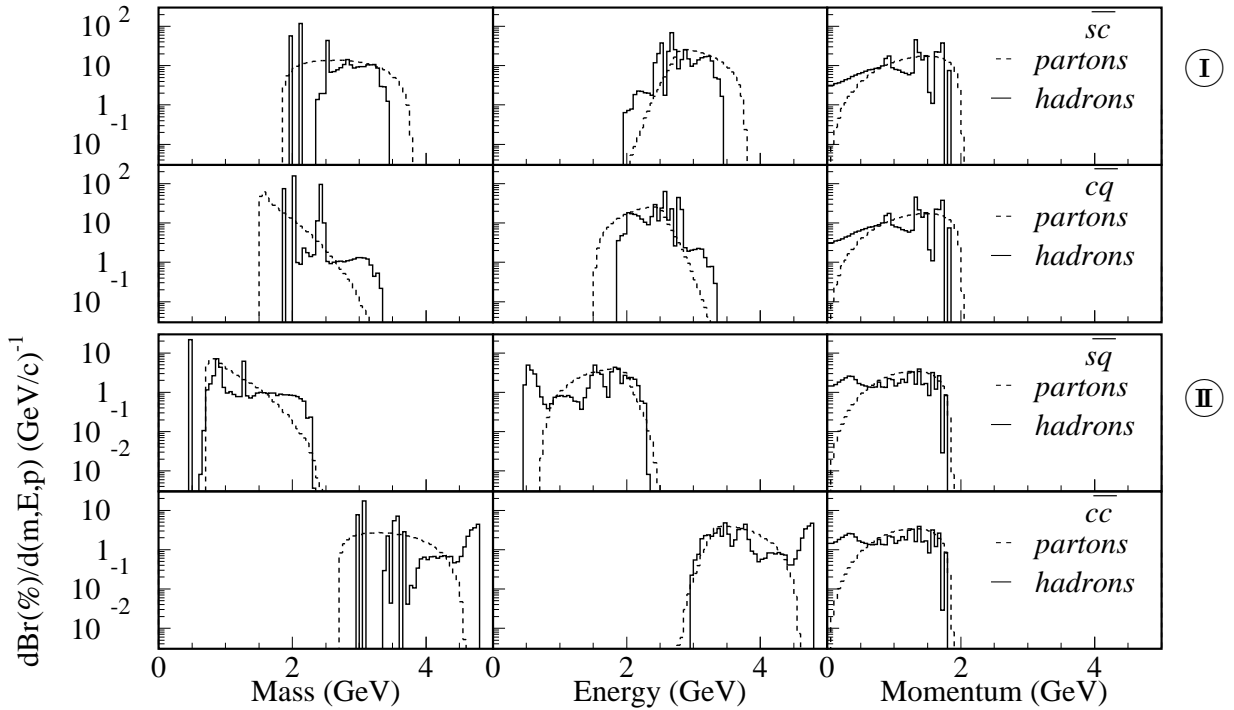}
\setlength{\unitlength}{10mm}
\begin{picture}(-0.2,7)
\linethickness{0.4mm}
\put(-0.2,8.8){\circle{0.5}}
\put(-0.6,8.4){\makebox(0.8,0.8){{$\mathbf{I}$}}}
\put(-0.2,4.4){\circle{0.5}}
\put(-0.6,4.0){\makebox(0.8,0.8){{$\mathbf{I\!I}$}}}
\end{picture}
\end{footnotesize}
\caption{Distributions of mass, energy and momentum for partons and final state
hadrons.
Dashed lines show the distributions of the initial partons.
Solid lines show the distributions of the final state particles originating
from the parton pair. $\bar{q}$ is either a $\bar{d}$ or a $\bar{u}$ quark
depending on if the meson is a $\bar{B}^0$ or a $B^-$.
Indicated in the figure are the color connections for the pair, see
Fig.~\ref{feynncrc}.}
\label{MEPdist}
\end{figure*}

Before comparing our model with data we want to emphasize the importance of the
model used for hadronisation. To illustrate this Fig.~\ref{MEPdist} shows the
mass, energy and momentum distributions for the parton string-systems obtained
from the matrix-element for each of the possible colour configurations compared
to the resulting  final state hadrons. As can be seen from the figure this
mapping is far from being smooth, especially when there is only one hadron
produced from the initial $q\bar{q}$ pair. It should also be noted that 
energy and momentum conservation implies that the energies of the 
$s\bar{c}$ and $c\bar{q}$ systems, and of $s\bar{q}$ and $c\bar{c}$, add up to the $B$-mass, as well as that their momentum distributions are pairwise the same.

\subsection{Normalization}
As specified in Eq.~(\ref{eqndiffdec}) we have included two different 
$K$-factors, one for semi-leptonic and one for hadronic decays, 
in order to
get a correct normalisation and to take into account the  
difference between semi-leptonic and hadronic decays.
We also be note that, since the $b$-quark mass is varying in the underlying 
$B$-meson model and the 
decay width is proportional to the $b$-mass to the fifth power, the $K$-factors are sensitive to the parameters of the $B$-meson model and not necessarily larger than 1.
The two $K$-factors have been obtained by simultaneously fitting the branching 
ratio for semileptonic decays $b \to e^- X$ to data using  $B^0/B^\pm
\to l^+ X= 10.24 \%$ \cite{PDG2008}, and also fitting the total
width from all the decays included in Table~\ref{Ktable} to the measured
lifetime. The resulting $K$-factors, for $p_F=0.57$ GeV, $m_{sp}=0.15$ GeV and assuming a total width of $4.20\times
10^{-13}$ GeV \cite{PDG2008}, are
\begin{equation}
K_{\rm sl} =0.75, \quad {K_{\rm h}}=1.4
\end{equation}
and the resulting $b$ branching ratios are shown given in Table \ref{Ktable}.
Of special interest here is the value for the $Br (b \to c \; s \;
\bar{c})$ (marked with boldface) which 
corresponds to the partonic mode that has been used to fit the model
and which is in good agreement with data, $Br (b \to c  s  \bar{c})=(22 \pm 4) \%$ \cite{PDG2008}.

\begin{table*} 
\begin{center}
\begin{ruledtabular}
\begin{tabular}{ccccc}
 & CKM- & Simulated decay & Branching ratio & Branching ratio \\ 
\raisebox{1.5ex}[0pt][0pt]{decay channel } & factors& width ($10^{-15}$ GeV) & without $K$-factors & with two $K$-factors \\ \hline 
$b  \to c \; e^-    \; \bar{\nu}_e    $  & $V_{cb} $            & \itshape    56.01 & \itshape    16.08\% & \itshape    10.14\% \\
$b  \to u \; e^-    \; \bar{\nu}_e    $  & $V_{ub} $            & \itshape     0.73 & \itshape     0.21\% & \itshape     0.13\% \\
$b  \to c \; \mu^-  \; \bar{\nu}_\mu  $  & $V_{cb} $            & \itshape    55.71 & \itshape    15.99\% & \itshape    10.08\% \\
$b  \to u \; \mu^-  \; \bar{\nu}_\mu  $  & $V_{ub} $            & \itshape     0.72 & \itshape     0.21\% & \itshape     0.13\% \\
$b  \to c \; \tau^- \; \bar{\nu}_\tau $  & $V_{cb} $            &    14.11 &     4.05\% &     2.55\% \\
$b  \to u \; \tau^- \; \bar{\nu}_\tau $  & $V_{ub} $            &     0.27 &     0.08\% &     0.05\% \\
\boldmath $b  \to c \; s      \; \bar{c}        $  & \boldmath $V_{cb} V_{cs}$      & \bfseries    63.40 & \bfseries    18.20\% & \bfseries    22.09\% \\
$b  \to c \; s      \; \bar{u}        $  & $V_{cb} V_{us}$      &     6.86 &     1.97\% &     2.39\% \\
$b  \to c \; d      \; \bar{c}        $  & $V_{cb} V_{cd}$      &     3.54 &     1.02\% &     1.23\% \\
$b  \to c \; d      \; \bar{u}        $  & $V_{cb} V_{ud}$      &   143.85 &    41.29\% &    50.11\% \\
$b  \to u \; s      \; \bar{c}        $  & $V_{ub} V_{cs}$      &     1.08 &     0.31\% &     0.38\% \\
$b  \to u \; s      \; \bar{u}        $  & $V_{ub} V_{us}$      &     0.09 &     0.03\% &     0.03\% \\
$b  \to u \; d      \; \bar{c}        $  & $V_{ub} V_{cd}$      &     0.06 &     0.02\% &     0.02\% \\
$b  \to u \; d      \; \bar{u}        $  & $V_{ub} V_{ud}$      &     1.92 &     0.55\% &     0.67\% \\
 All channels                                  &                      &   348.35 &     100.\% &     100.\% \\ \hline
 Total fraction of c                           &                      &          &     118.\% &     122.\% \\
\end{tabular}
\end{ruledtabular}
\end{center}
\caption{Decay widths and branching ratios for $b$-quark decays used to fix the
$K$-factors. The semileptonic decays, marked with italic, and the total 
width are used to fix the factors $K_{\rm sl}$ and $K_{\rm h}$. The channel $b \to c \; s \; \bar{c}$ is used to normalize our results.}
\label{Ktable}
\end{table*}

\subsection{Summary of the model}
The complete new model contains 4 parameters:
\begin{itemize}
\item the width $p_F$ of the Gaussian $b$-quark momentum distribution in the $B$-meson,  
\item the mass $m_{sp}$ of the spectator quark in the $B$-meson 
\item the probability $P$ for soft colour exchange, 
\item the probability $P_{\rm AVM}$ for producing a meson with $L=1, S=0, J=1$
\end{itemize}
The parameter values given in Table \ref{resparam} are determined by fitting to $B$-meson decay data. The values of $p_F$ and $m_{sp}$ of the ACCMM-model are sensitive to the momentum spectrum of the produced $J/\psi$, whereas $P$ and $P_{\rm AVM}$ are fitted to the branching ratios. This is done by simulating a large number of $B$ decays where 
$b \to c \; s \; \bar{c}$ and then calculating the
branching ratios using the normalization described above.
A $\chi^2$ is then calculated as
\begin{equation}
\chi^2=\sum\frac{\left(BR_{exp}-BR_{data}\right)^2}{\sigma_{BR_{exp}}^2}\;.
\end{equation}
Two sets of branching ratios have been used for the fits, one with only
inclusive branching ratios and one with both inclusive and exclusive ones. When
changing $p_F$ and $m_{sp}$ 
within the ranges given in Table~\ref{resparam}
it is always possible to find a reasonable fit to the branching ratios. We
have tried a number of different combinations of parameter
values, including the ones used in~\cite{Palmer:1996dy,NRQCD-mom}
and the one that gives the best fit to
the direct $J/\psi$ momentum distribution is then used. 

As discussed in Section \ref{smallen}, decays to mesons with $L=1$ (e.g.\ $K_1(1270)$) requires to account for production of the much heavier axial vector mesons and we obtain $P_{\rm AVM}\approx 0.7$ for the probability that an $S=0$ meson is in an $L=1$ state. We note that this value is close to 3/4 as obtained by simple counting of available angular momentum states for $L=0$ and 1. (In default \pythia\ $P_{\rm AVM}=0$ since observables based on 'stable' hadrons are not sensitive to whether these mesons have been produced as intermediate states or not.) The inclusion of such heavier mesons requires a retuning of the \pythia\ parameter $n_{try}$ controlling the number of tries allowed in particle formation from small mass systems. By increasing from the default value $n_{try}=2$ one accounts better for the available phase space, and we have chosen $n_{try}=6$ as preferred when fitting to all branching ratios. Fitting only inclusive branching ratios would prefer a slightly higher value, $n_{try}=9$, but this degrades the fit to all branching ratios more than using $n_{try}=6$ does for the inclusive fit and, in addition, gives a worse fit to the $J/\psi$ momentum distribution.

\begin{table}
\begin{center}
\begin{ruledtabular}
\begin{tabular}{ccccc}
 & & \multicolumn{3}{c}{Fitted to} \\ 
Parameter & Range & inclusive BR & all BR & $J/\psi$ momentum \\ \hline
$p_F$     & 0.30 -- 0.60 & & & 0.57 GeV\\ 
$m_{sp}$  & 0.05 -- 0.15 & & & 0.15 GeV\\ 
$P$       & 0 -- 1 & 0.16 & 0.16 & \\ 
$P_{\rm AVM}$ & 0 -- 1 & 0.71 & 0.72 & \\ 
\end{tabular}
\end{ruledtabular}
\end{center}
\caption{The parameters of the model: 
the width $p_F$ of the Gaussian $b$-quark momentum distribution in the $B$-meson,  
the mass $m_{sp}$ of the spectator quark in the $B$-meson 
the probability $P$ for soft colour exchange, 
the probability $P_{\rm AVM}$ for producing a meson with $L=1, S=0, J=1$. 
Their values are obtained by fitting to: 
the inclusive branching ratios in Fig.~\ref{plotBrinc},
 all branching ratios, i.e.\ also including the exclusive ones in
Fig.~\ref{plotBrexc}, and the $J/\psi$ momentum spectrum in
Fig.~\ref{Momdist}.}
\label{resparam}
\end{table}

\subsection{Comparison to data}
\subsubsection{Inclusive branching ratios}

\begin{figure}
\includegraphics*[width=7cm]{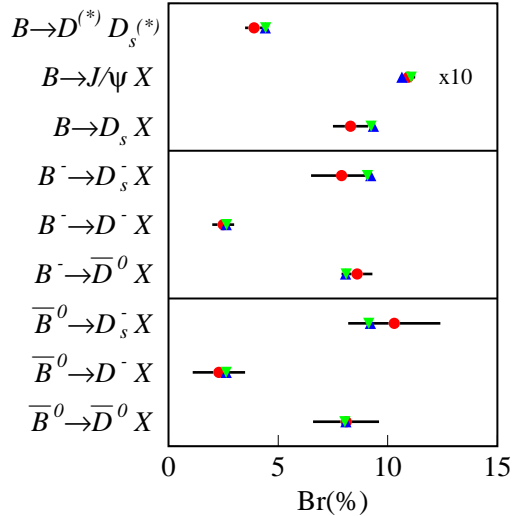}
\caption{Comparison between data \cite{PDG2008} (red dots with error bars)
of specified inclusive decay channels for $B$-mesons and the model fitted to
these data only (green down-pointing triangles) and to all branching ratios
(blue up-pointing triangles). The lower part is for $\bar{B}^0$, the middle part
for $B^-$ and the top part for $\bar{B}^0$ and $B^-$ combined. 
Note that the $B\to J/\psi X$
channel is multiplied with a factor 10 to make it more visible.}
\label{plotBrinc}
\end{figure}

Fig.~\ref{plotBrinc} shows the results of the fit to the inclusive branching ratios as well as the fit to all branching ratios compared to data. As is clear from the figure, in both cases the model describes the inclusive
branching ratios very well. In fact, when fitting to these inclusive channels only, we
obtain 
\begin{equation}
\frac{\chi^2}{d.o.f.}=\frac{5.0}{7}=0.7 \;.
\end{equation}
We also note that the difference between the two fits is very small, 
it is only barely visible for the $B \to J/\psi X$ and $B \to  D_s X$ channels.

\subsubsection{Exclusive branching ratios}

\begin{figure*}
\begin{footnotesize}
\setlength{\unitlength}{10mm}
\includegraphics*[width=7cm]{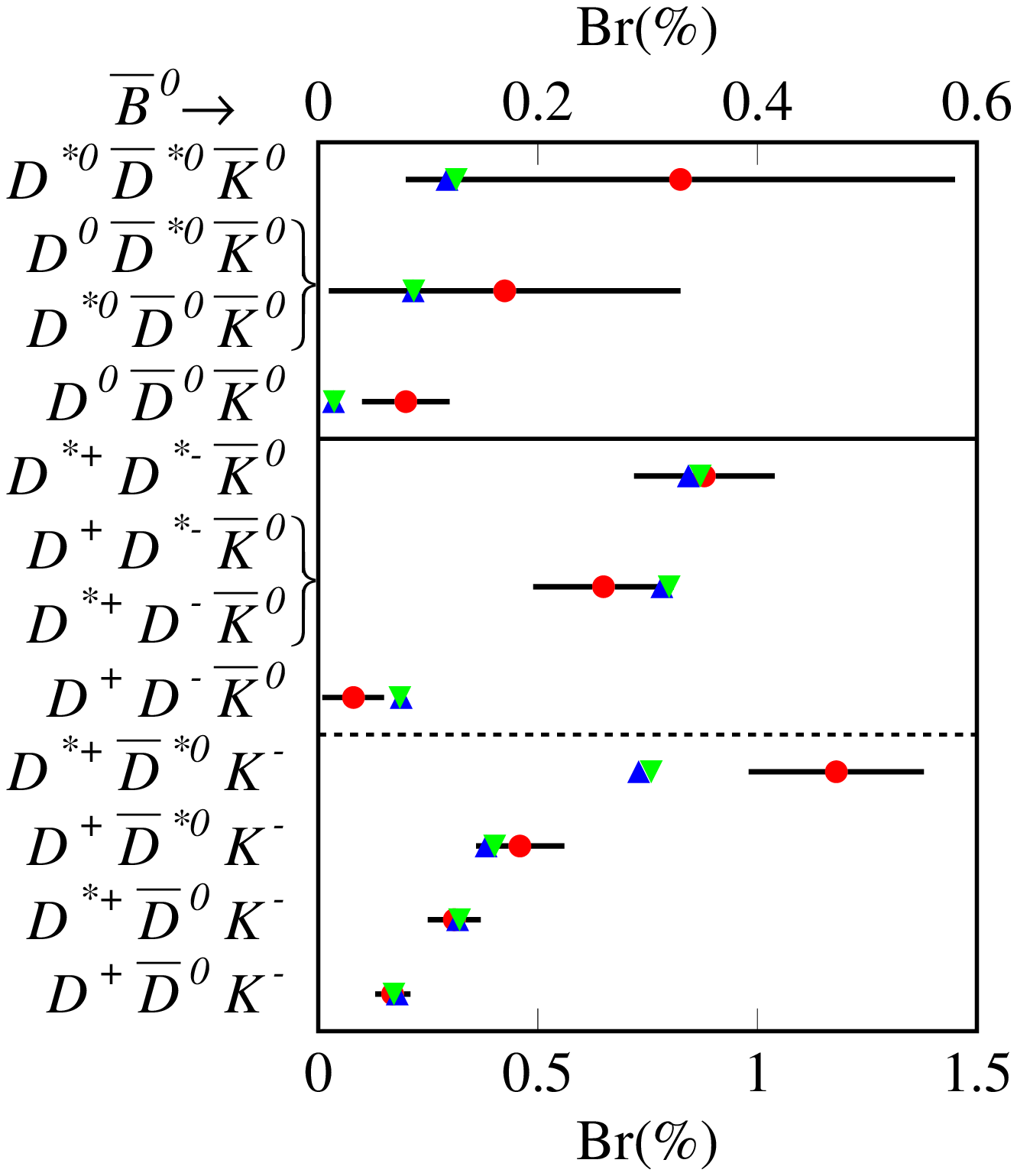}
\begin{picture}(-0.2,7)
\linethickness{0.4mm}
\put(-1.1,6.1){\makebox(0.8,0.8){a}}
\put(-1.1,5.9){\circle{0.5}}
\put(-1.5,5.5){\makebox(0.8,0.8){{$\mathbf{I\!I}$}}}
\put(-1.7,3.9){\circle{0.5}}
\put(-2.1,3.5){\makebox(0.8,0.8){{$\mathbf{I}$}}}
\put(-1.7,3.5){\makebox(0.8,0.8){{$\mathbf{+}$}}}
\put(-0.9,3.9){\circle{0.5}}
\put(-1.3,3.5){\makebox(0.8,0.8){{$\mathbf{I\!I}$}}}
\put(-1.1,1.9){\circle{0.5}}
\put(-1.5,1.5){\makebox(0.8,0.8){{$\mathbf{I}$}}}
\end{picture}
\includegraphics*[width=7cm]{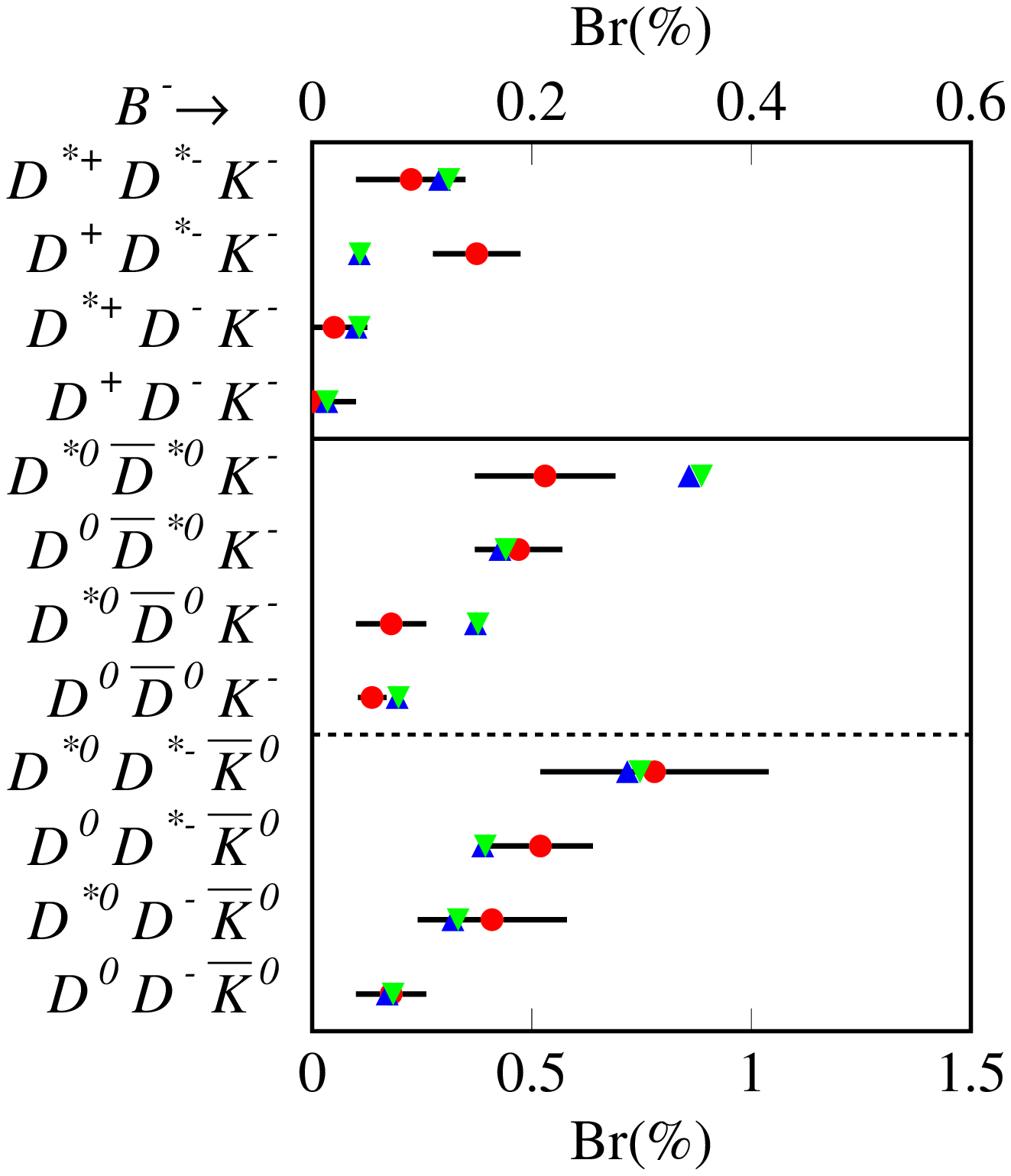}
\begin{picture}(-0.2,7)
\linethickness{0.4mm}
\put(-1.1,6.1){\makebox(0.8,0.8){b}}
\put(-1.1,5.9){\circle{0.5}}
\put(-1.5,5.5){\makebox(0.8,0.8){{$\mathbf{I\!I}$}}}
\put(-1.7,3.9){\circle{0.5}}
\put(-2.1,3.5){\makebox(0.8,0.8){{$\mathbf{I}$}}}
\put(-1.7,3.5){\makebox(0.8,0.8){{$\mathbf{+}$}}}
\put(-0.9,3.9){\circle{0.5}}
\put(-1.3,3.5){\makebox(0.8,0.8){{$\mathbf{I\!I}$}}}
\put(-1.1,1.9){\circle{0.5}}
\put(-1.5,1.5){\makebox(0.8,0.8){{$\mathbf{I}$}}}
\end{picture}

\includegraphics*[width=7cm]{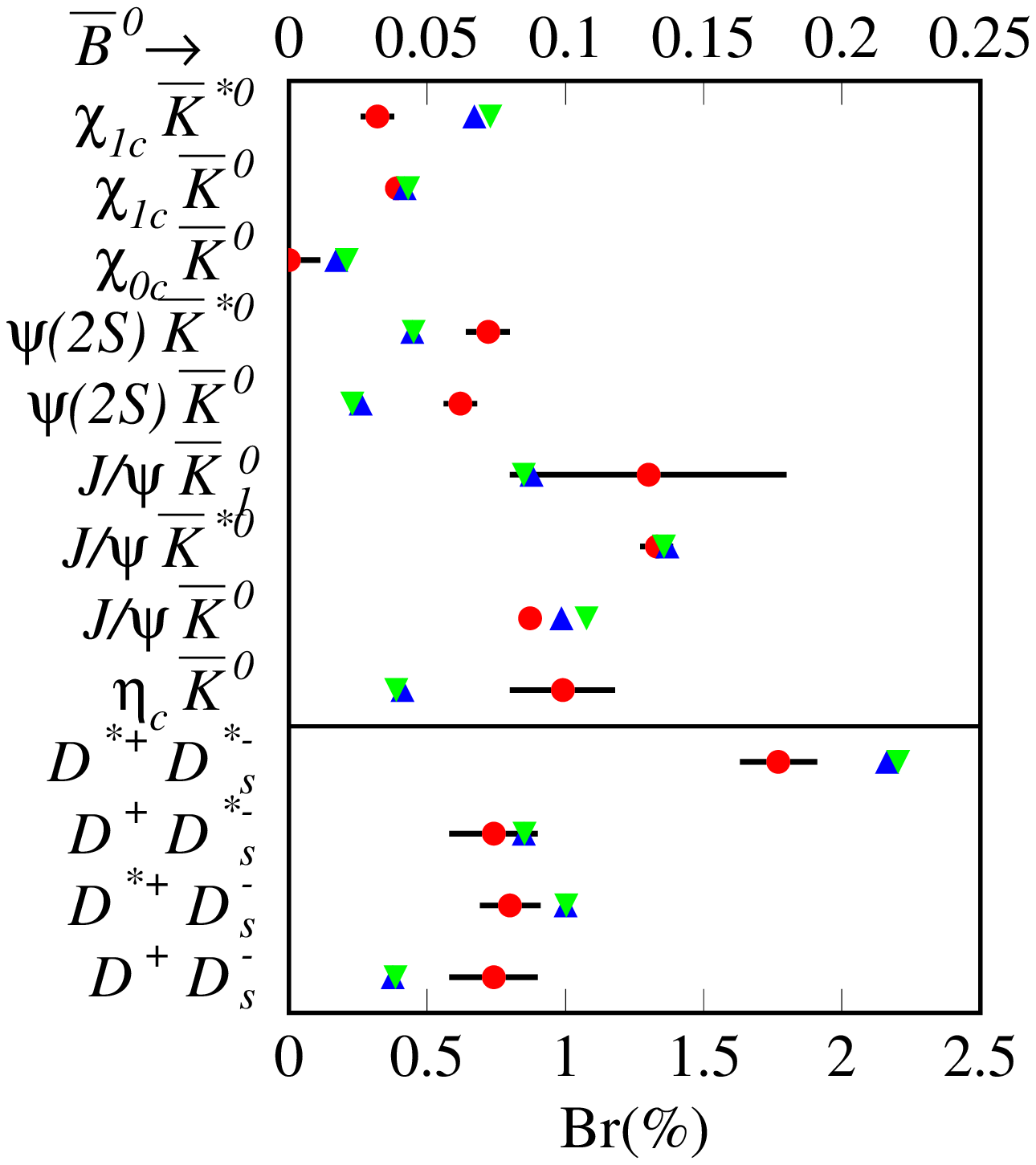}
\begin{picture}(-0.2,7)
\linethickness{0.4mm}
\put(-1.1,6.1){\makebox(0.8,0.8){c}}
\put(-1.1,5.9){\circle{0.5}}
\put(-1.5,5.5){\makebox(0.8,0.8){{$\mathbf{I\!I}$}}}
\put(-1.1,1.9){\circle{0.5}}
\put(-1.5,1.5){\makebox(0.8,0.8){{$\mathbf{I}$}}}
\end{picture}
\includegraphics*[width=7cm]{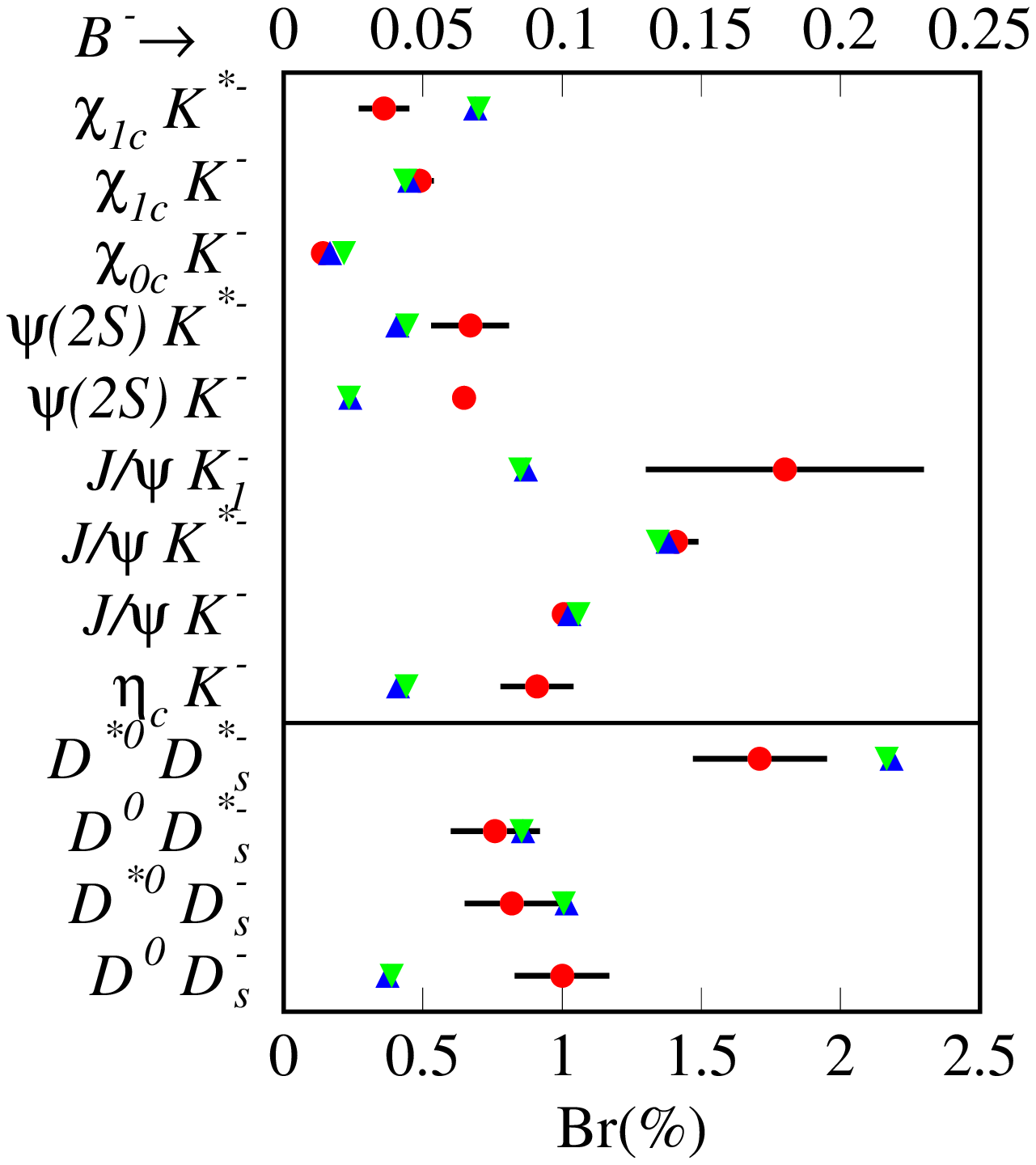}
\begin{picture}(-0.2,7)
\linethickness{0.4mm}
\put(-1.1,6.1){\makebox(0.8,0.8){d}}
\put(-1.1,5.9){\circle{0.5}}
\put(-1.5,5.5){\makebox(0.8,0.8){{$\mathbf{I\!I}$}}}
\put(-1.1,1.9){\circle{0.5}}
\put(-1.5,1.5){\makebox(0.8,0.8){{$\mathbf{I}$}}}
\end{picture}
\end{footnotesize}
\caption{
Comparison between data (red dots with error bars from \cite{BaBardata} in (a,b)
and from \cite{PDG2008} in (c,d)) of specified exclusive decay channels for
$B^0$ (a,c) and $B^-$ (b,d) and the model fitted to only the inclusive
decay channels in Fig.~\ref{plotBrinc}  (green down-pointing triangles) and also
including these exclusive channels (blue up-pointing triangles). 
The lower parts are associated with the original colour string configuration
$\mathbf{I}$, whereas the upper parts (with upper BR scale) are for
configuration $\mathbf{I\!I}$ (cf.\ Fig.~\ref{feynncrc}). In (a,b) there are
middle parts where both $\mathbf{I}$ and $\mathbf{I\!I}$ contribute.
}
\label{plotBrexc}
\end{figure*}

Fig.~\ref{plotBrexc} demonstrates that also the exclusive branching ratios are
quite well described by the model and that the difference between the two fits is again very small. In both cases the model essentially describes all the different channels involving
$D$-mesons, including the relative strength of $D$ and $D^{\ast}$ channels.
It also gives an overall good description of the states that can be produced from the two colour configurations displayed in Fig.~\ref{feynncrc} and discussed in Section~\ref{sec:colstr} respectively. This is true both for the two-body decays shown  in Fig.~\ref{plotBrexc}(c,d)  as well as for the three-body decays  shown  in Fig.~\ref{plotBrexc}(a,b) where for some channels both colour configuration  $\mathbf{I}$ and $\mathbf{I\!I}$ contribute as indicated in the figure. As a consequence the model also gives sum-rules of the type, $Br(\bar{B}^0 \to D^{(\ast)+}D^{(\ast)-}\bar{K}^0) \simeq
Br(\bar{B}^0 \to D^{(\ast)+}\bar{D}^{(\ast)0}{K}^-)+
Br(\bar{B}^0 \to D^{(\ast)0}\bar{D}^{(\ast)0}\bar{K}^0)$, which, at least within errors, are in agreement with data.
Evidently it is more demanding to describe all these
exclusive final states which are sensitive to the non-perturbative dynamics in
hadronic few-body systems with relatively small kinetic energy available. In
view of this, it is remarkable that this model with only a few parameters
reproduce the data so well. 
We also note that the rate for $J/\psi K_1(1270)$, which is controlled by the $P_{\rm AVM}$ parameter, comes out essentially right even if this decay mode is not included in the fit.

 Not surprisingly, however, there are some channels
where the description is not so good.
This is mainly for channels with one of the heavier charmonium states and  
either a $K$ or a $K^{\ast}$, e.g. $\bar{B}^0\to \chi_{1c}
\bar{K}^{\ast 0}$. In our approach the two mesons are formed more or less
independently and we therefore get about the same ratio for $\bar{K}^0$ and
$\bar{K}^{\ast 0}$ irrespectively of whether it is produced together with a heavier or lighter charmonium state. However, the data indicates that the production of mesons with small relative momenta, {\it i.e.} closer to threshold, such as $\chi_{1c}
\bar{K}^{\ast 0}$, should be suppressed.

When fitting to all channels we get $\chi^2/d.o.f.=353/46=7.7$, which may be
considered too large for a good fit. Most of the $\chi^2$ is, however, coming
from some particular channels. For example, removing the $B\to \psi (2S) K$ channels,
which gives the dominant contribution, results in $\chi^2/d.o.f.=181/44=4.1$.
We also note that since our model is not based on first principles, it is not very meaningful to
perform $\chi^2$-tests of it. Instead, it is meant to investigate whether the
SCI model, which is phenomenologically very successful in describing other kinds
of data related to colour string-field topologies, is of relevance also in
decays of $B$-mesons, and this seems indeed to be the case.

\subsubsection{The $J/\psi$ momentum distribution}

\begin{figure*}
\includegraphics*[width=16cm]{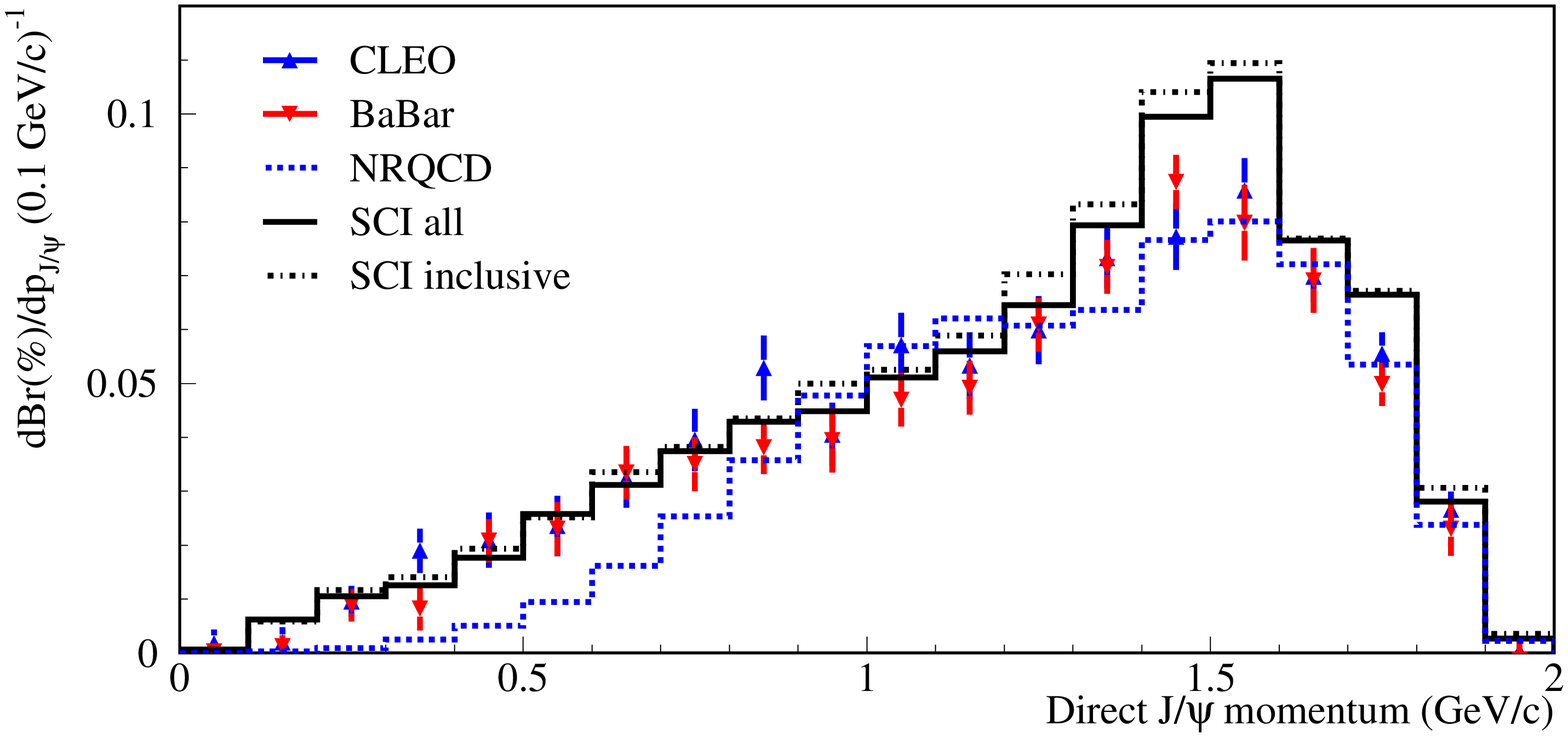}
\caption{
Momentum distribution of directly produced $J/\psi$ in $B\to J/\psi X$ decays in the $\Upsilon$(4S) rest frame. Data from \cite{CLEOJpsi} and \cite{BaBarJpsi} compared to different models: our soft colour interaction model fitted to only inclusive decay branching ratios and to all branching ratios, and 
the NRQCD curve which is a combination of the COM \cite{NRQCD-mom} and $J/\psi K(^{\ast})$ done in \cite{BaBarJpsi}.
} 
\label{Momdist}
\end{figure*}

Finally, in  Fig.~\ref{Momdist} the results of our model is compared to the 
momentum distribution of directly produced $J/\psi$'s in $B$ decays measured in the $\Upsilon$(4S) rest frame. First of all it is clear from the figure that our model gives a good overall description of the data especially in the low-momentum region, whereas it is slightly too high in the peak region. At the same time it should be noted that the model has not been fitted to the normalisation of these data, only to the inclusive and inclusive plus exclusive branching ratios respectively. Since the peak in Fig.~\ref{Momdist} is dominated by two-body decays, the model's excess here of $\sim 20\%$ indicates that it produces somewhat too many direct $J/\psi$'s, in particular in the $J/\psi K$ channel as is also indicated by figure Fig.~\ref{plotBrexc}(c).

For comparison  Fig.~\ref{Momdist} also shows the results of a NRQCD based model from~\cite{BaBarJpsi} discussed in Section 2. This model is a combination of the COM  results~\cite{NRQCD-mom} together with a model for the exclusive  $J/\psi K$  and $J/\psi K^{\ast}$ decays. In contrast to our model, the NRQCD based model is not at all able to describe the low-momentum region, which has given rise to various alternative explanations as already mentioned. 
It is also important to recognize that the parameter that mostly affects the overall shape of the momentum distribution, namely the width, $p_F$, of the Gaussian momentum distribution  of the $b$-quark in the $B$-meson, does not affect the low-momentum region of the spectrum. Instead this parameter is responsible for the smearing in the peak region. Similarly, the mass $m_{sp}$ of the spectator quark also does not affect the low-momentum region. In addition, for both of these parameters we have used the same values as in~\cite{Palmer:1996dy}. The main difference of our model compared to the NRQCD-based ones therefore lies in the dynamical treatment of the decay products from the $b$-quark.
Such soft dynamics is particularly important to get a good description of the low-momentum region in Fig.~\ref{Momdist}, and our detailed model does indeed provide an improvement.

\section{Conclusions}

In this paper we have presented a new model, which can describe both open and hidden charm production in $B$-meson decays. The model is based on the ACCMM model for the internal $B$-meson dynamics, the SCI model for colour rearrangements in the partonic final state, a model for mapping of colour singlet $c \bar{c}$ pair on to charmonium states and a new procedure for hadronising colour singlet systems with small invariant mass within the Lund string fragmentation framework. 

Using more or less standard values for the parameters of the ACCMM and SCI models:
the width of the Gaussian momentum distribution  of the $b$-quark in the $B$-meson, $p_F=0.57$ GeV, the mass of the spectator quark, $m_{sp}=0.15$ GeV, and, the probability for a colour rearrangement, $P=0.16$, we find overall good agreement with the data both on open and hidden charm production. In order for this to be possible we have had to activate the production of axial vector mesons in the hadronisation with a probability $P_{\rm AVM}=0.7$ consistent with the number of available angular momentum states. We have also improved the probing of the available phase space, which is particularly constrained for processes close to mass thresholds, by increasing the number of times that the program tries to make a single hadron out of a given small mass partonic system. Related to this, we have modified the way that energy and momentum is exchanged when the invariant mass of such a parton system has to be changed in order to give the proper hadron mass.

Our model gives a very good description of inclusive observables such as $B \to J/\psi X$ and $B \to D_s X$, which shows that the basic ideas ingredients of the model are correct. In particular it shows that the idea of soft colour interactions also can be successfully applied in $B$-meson decays giving a unified description both of open and hidden charm production. When it comes to exclusive decay modes the overall description is still good but there are some channels, which are not well described. The latter is especially true for $B \to \psi(\mbox{2S})K^{(\ast)}$ and
$B \to \chi_{1c} K^{\ast}$ indicating that there is something lacking in the model for mapping colour singlet $c \bar{c}$ pair onto charmonium states which also may be connected to the fact that these decay modes are the ones closest to threshold that we have considered. Last but not least, the model also describes the momentum distribution of direct $J/\psi$'s, including the low mass region where earlier models, most notably the COM, fails. This shows that there is no need to invoke new hadronic states such as hybrids or bound diquark anti-diquark states to explain this region. Instead this comes out naturally from our model as a consequence of the non-perturbative dynamics involved in the hadronisation process.

The overall conclusion of our study is that it is possible to describe and understand a wealth of data on $B$-meson decays with our relatively simple model based on the framework of Soft Colour Interactions, which modify the colour structure of an event and thereby the string-field topology leading to different hadronic final states. The fact that this SCI model has earlier been successful in describing a wide range of phenomena, such as rapidity gap events and charmonium production in both hadron-hadron and electron-proton collisions, and now $B$-meson decays shows that it captures essential {\em generic} features of non-perturbative QCD interactions.


\begin{thebibliography}{0}
\expandafter\ifx\csname natexlab\endcsname\relax\def\natexlab#1{#1}\fi
\expandafter\ifx\csname bibnamefont\endcsname\relax
  \def\bibnamefont#1{#1}\fi
\expandafter\ifx\csname bibfnamefont\endcsname\relax
  \def\bibfnamefont#1{#1}\fi
\expandafter\ifx\csname citenamefont\endcsname\relax
  \def\citenamefont#1{#1}\fi
\expandafter\ifx\csname url\endcsname\relax
  \def\url#1{\texttt{#1}}\fi
\expandafter\ifx\csname urlprefix\endcsname\relax\def\urlprefix{URL }\fi
\providecommand{\bibinfo}[2]{#2}
\providecommand{\eprint}[2][]{\url{#2}}

\end{thebibliography}


\begin{thebibliography}{99}
\addcontentsline{toc}{section}{\numberline{}References}

\bibitem{Lundstring}
B.~Andersson, G.~Gustafson, G.~Ingelman and T.~Sj\"ostrand,
Phys.\ Rept.\  {\bf 97} (1983) 31.

\bibitem{Ingelman:1984ns}
  G.~Ingelman and P.~E.~Schlein,
  Phys.\ Lett.\  B {\bf 152} (1985) 256.

\bibitem{SCI}
A.~Edin, G.~Ingelman and J.~Rathsman,
Phys.\ Lett.\ B {\bf 366} (1996) 371;\\
%
A.~Edin, G.~Ingelman and J.~Rathsman,
Z.\ Phys.\ C {\bf 75} (1997) 57.

\bibitem{Ingelman:1996mq}
  G.~Ingelman, A.~Edin and J.~Rathsman,
  Comput.\ Phys.\ Commun.\  {\bf 101}, 108 (1997)
  [arXiv:hep-ph/9605286].

\bibitem{pythia}
T.~Sj\"ostrand, P.~Eden, C.~Friberg, L.~Lonnblad, G.~Miu, S.~Mrenna and E.~Norrbin,
Comput.\ Phys.\ Commun.\  {\bf 135} (2001) 238
[arXiv:hep-ph/0010017].

\bibitem{Enberg:2001vq}
  R.~Enberg, G.~Ingelman and N.~Timneanu,
  Phys.\ Rev.\  D {\bf 64} (2001) 114015
  [arXiv:hep-ph/0106246].

\bibitem{Edin:1997zb}
  A.~Edin, G.~Ingelman and J.~Rathsman,
  Phys.\ Rev.\  D {\bf 56} (1997) 7317
  [arXiv:hep-ph/9705311].

\bibitem{BrennerMariotto:2001sv}
  C.~Brenner Mariotto, M.~B.~Gay Ducati and G.~Ingelman,
  Eur.\ Phys.\ J.\  C {\bf 23}, 527 (2002)
  [arXiv:hep-ph/0111379].

\bibitem{Ingelman:2005ku}
  G.~Ingelman,
  Int.\ J.\ Mod.\ Phys.\  A {\bf 21}, 1805 (2006)
  [arXiv:hep-ph/0512146].

\bibitem{Brodsky:2002ue}
  S.~J.~Brodsky, P.~Hoyer, N.~Marchal, S.~Peigne and F.~Sannino,
  Phys.\ Rev.\  D {\bf 65}, 114025 (2002)
  [arXiv:hep-ph/0104291].

\bibitem{Brodsky:2004hi}
  S.~J.~Brodsky, R.~Enberg, P.~Hoyer and G.~Ingelman,
  Phys.\ Rev.\  D {\bf 71} (2005) 074020
  [arXiv:hep-ph/0409119].

\bibitem{GAL}
J.~Rathsman,
Phys.\ Lett.\ B {\bf 452} (1999) 364.

\bibitem{Zapp:2005kt}
  K.~Zapp, G.~Ingelman, J.~Rathsman and J.~Stachel,
  Phys.\ Lett.\  B {\bf 637} (2006) 179
  [arXiv:hep-ph/0512300].

\bibitem{DeGrand:1979wf}
  T.~A.~DeGrand and D.~Toussaint,
  Phys.\ Lett.\  B {\bf 89} (1980) 256.

\bibitem{Wise:1979tp}
  M.~B.~Wise,
  Phys.\ Lett.\  B {\bf 89} (1980) 229.

\bibitem{Beneke:1998ks}
  M.~Beneke, F.~Maltoni and I.~Z.~Rothstein,
  Phys.\ Rev.\  D {\bf 59} (1999) 054003
  [arXiv:hep-ph/9808360].

\bibitem{Bergstrom:1994vc}
  L.~Bergstrom and P.~Ernstrom,
  Phys.\ Lett.\  B {\bf 328} (1994) 153
  [arXiv:hep-ph/9402325].

\bibitem{NRQCD}
G.~T.~Bodwin, E.~Braaten and G.~P.~Lepage,
Phys.\ Rev.\ D {\bf 51} (1995) 1125
[Erratum-ibid.\ D {\bf 55} (1997) 5853]
[arXiv:hep-ph/9407339].

\bibitem{Ko:1995iv}
  P.~Ko, J.~Lee and H.~S.~Song,
  Phys.\ Rev.\  D {\bf 53} (1996) 1409
  [arXiv:hep-ph/9510202].

\bibitem{Palmer:1996dy}
  W.~F.~Palmer, E.~A.~Paschos and P.~H.~Soldan,
  arXiv:hep-ph/9602376.

\bibitem{NRQCD-mom}
M.~Beneke, G.~A.~Schuler and S.~Wolf,
Phys.\ Rev.\ D {\bf 62} (2000) 034004
[arXiv:hep-ph/0001062].

\bibitem{BaBarJpsi}
B.~Aubert {\it et al.}  [BABAR Collaboration],
Phys.\ Rev.\ D {\bf 67} (2003) 032002
[arXiv:hep-ex/0207097].

\bibitem{Brodsky:1997yr}
  S.~J.~Brodsky and F.~S.~Navarra,
  Phys.\ Lett.\  B {\bf 411} (1997) 152
  [arXiv:hep-ph/9704348].

\bibitem{Close:2003ae}
  F.~E.~Close and J.~J.~Dudek,
  Phys.\ Rev.\  D {\bf 69} (2004) 034010
  [arXiv:hep-ph/0308098].

\bibitem{Chua:2003fp}
  C.~K.~Chua, W.~S.~Hou and G.~G.~Wong,
  Phys.\ Rev.\  D {\bf 68} (2003) 054012
  [arXiv:hep-ph/0305180].

\bibitem{Bigi:2005fr}
  I.~Bigi, L.~Maiani, F.~Piccinini, A.~D.~Polosa and V.~Riquer,
  Phys.\ Rev.\  D {\bf 72} (2005) 114016
  [arXiv:hep-ph/0510307].

\bibitem{Bauer:1986bm}
  M.~Bauer, B.~Stech and M.~Wirbel,
  Z.\ Phys.\  C {\bf 34} (1987) 103.

\bibitem{Rosner:1990xx}
  J.~L.~Rosner,
  Phys.\ Rev.\  D {\bf 42} (1990) 3732.

\bibitem{Luo:2001mc}
  Z.~Luo and J.~L.~Rosner,
  Phys.\ Rev.\  D {\bf 64} (2001) 094001
  [arXiv:hep-ph/0101089].

\bibitem{Chen:2005rp}
  C.~H.~Chen, C.~Q.~Geng and Z.~T.~Wei,
  Eur.\ Phys.\ J.\  C {\bf 46} (2006) 367
  [arXiv:hep-ph/0507295].

 \bibitem{Thomas:2005bu}
  C.~E.~Thomas,
  Phys.\ Rev.\  D {\bf 73} (2006) 054016
  [arXiv:hep-ph/0511169].

\bibitem{Isgur:1988gb}
  N.~Isgur, D.~Scora, B.~Grinstein and M.~B.~Wise,
  Phys.\ Rev.\  D {\bf 39} (1989) 799.

\bibitem{Colangelo:2002dg}
  P.~Colangelo and F.~De Fazio,
  Phys.\ Lett.\  B {\bf 532} (2002) 193
  [arXiv:hep-ph/0201305].

 \bibitem{Datta:2003re}
  A.~Datta and P.~J.~O'donnell,
  Phys.\ Lett.\  B {\bf 572} (2003) 164
  [arXiv:hep-ph/0307106].

\bibitem{ACCMM}
G.~Altarelli, N.~Cabibbo, G.~Corbo, L.~Maiani and G.~Martinelli,
Nucl.\ Phys.\ B {\bf 208} (1982) 365.

\bibitem{Collider}
V.~D.~Barger and R.~J.~Phillips,
\emph{Collider Physics},
Addison-Wesley, 1987

\bibitem{PDG2008}
  C.~Amsler {\it et al.}  [Particle Data Group],
  Phys.\ Lett.\  B {\bf 667} (2008) 1.

\bibitem{BaBardata}
B.~Aubert  [BaBar Collaboration],
arXiv:hep-ex/0305003.

\bibitem{CLEOJpsi}
R.~Balest {\it et al.}  [CLEO Collaboration],
Phys.\ Rev.\ D {\bf 52} (1995) 2661.
S.~Anderson {\it et al.}  [CLEO Collaboration],
arXiv:hep-ex/0207059.

\end{thebibliography}
\end{document}